# Analytical solution for inviscid flow inside an evaporating sessile drop

Hassan Masoud and James D. Felske

*Department of Mechanical and Aerospace Engineering,*

*State University of New York at Buffalo,*

*Buffalo, New York 14260, USA*

Inviscid flow within an evaporating sessile drop is analyzed. The field equation, $E^2\psi = 0$, is solved for the stream function. The exact analytical solution is obtained for arbitrary contact angle and distribution of evaporative flux along the free boundary. Specific results and computations are presented for evaporation corresponding to both uniform flux and purely diffusive gas phase transport into an infinite ambient. Wetting and non-wetting contact angles are considered with flow patterns in each case being illustrated. The limiting behaviors of small contact angle and droplets of hemispherical shape are treated. All of the above categories are considered for the cases of droplets whose contact lines are either pinned or free to move during evaporation.

47.55.D-

## I. INTRODUCTION

Recently, the problem of sessile drop evaporation has found prominence in relation to the deposition of particles which occurs during the drying of colloidal drops. Deposition in ring patterns (the 'coffee ring effect') influences a variety of applications: DNA mapping [1,2], ink-jet printing [3,4,5], production of crystals [6,7,8], coating with paint. Apart from ring patterns, applications include: buckling instability and skin formation by deposition from polymer solutions [9-11], and evaporation of liquid drops to cool a hot surface [12].

Understanding these phenomena requires the determination of the distribution of evaporative flux along the free-interface and the velocity distribution engendered in the liquid phase due to this evaporation. Depending on the interplay of these two factors, the direction of the free-surface flow can be either towards or away from the contact line.

A number of investigations have previously focused on this problem. Popov [13] solved exactly for the vapor phase transport from which the distribution of evaporative flux is determined. A useful approximate representation of this result was subsequently developed by Hu & Larson [14] for contact angles less than ninety degrees. An integral analysis was presented by Deegan [15-17] and Popov [13] for determining the radial distribution of the vertically averaged radial velocity. This result was used to analyze the limit of small contact angle. A numerical solution was pursued by Fischer [18] for viscous flow in the lubrication theory limit. Other numerical efforts have been presented by Hu & Larson [19] and Widjaja et al. [20] who solved the Stokes flow for contact angles ranging from zero to ninety degrees. A semi-analytical solution was obtained by Hu & Larson [19] in the lubrication theory limit. Tarasevich [21] and Petsi and Burganos [22,23] derived exact analytical solutions for irrotational flow within hemispheres, hemicylinders and cylindrical caps respectively.

The focus of the present work is on deriving the exact analytical solution for the irrotational flow within axisymmetric evaporating drops of arbitrary contact angle ($0 \leq \theta_c \leq \pi$) and evaporative flux distribution along the free surface. The behavior is considered within the context of the contact line of the droplet being either pinned or free to move during evaporation. Analytical expressions for the expansion coefficient are given for the limiting cases of droplets having either small contact angle or being hemispherical in shape.

## II. GEOMETRY, MODEL, AND SOLUTION

### A. Geometry

A sessile drop has an equilibrium shape given by energy minimization analysis [24]. For small drops, the influence of gravity on the shape is insignificant and the drop takes the shape of a spherical-cap. The boundary of the spherical-cap is exactly mapped in toroidal coordinates – see Fig. 1. These are therefore the natural coordinates to adopt in mathematically analyzing the phenomena.

Flow within the drop is independent of azimuthal angle ($\equiv$ 3-D, axisymmetric). Shown in Fig. 1 is a cut through the toroidal geometry at a given azimuthal angle, $\varphi$. The cross-sectional toroidal coordinates ($\alpha, \theta$) are indicated in the figure along with cylindrical coordinates ($r, z$) where, in relation to Cartesian coordinates, $r = (x^2 + y^2)^{1/2}$. The angle $\theta$, measured in the same sense as the contact angle $\theta_c$, is related to the angle $\beta$ used in [25] by: $\theta = \pi - \beta$. The metric coefficients for the toroidal geometry [25], when written in terms of $\theta$, are then

$$h_\alpha = h_\theta = h_\phi / \sinh\alpha = R(\cosh\alpha + \cos\theta)^{-1}, \tag{1}$$

in which $0 \leq \alpha \leq \infty$, $-\pi \leq \theta \leq \pi$, $0 \leq \varphi \leq 2\pi$, and $R$ is the distance from the z-axis to the contact line. The relationships between toroidal and cylindrical coordinates are:

$$r / \sinh\alpha = z / \sin\theta = R(\cosh\alpha + \cos\theta)^{-1}, \tag{2}$$

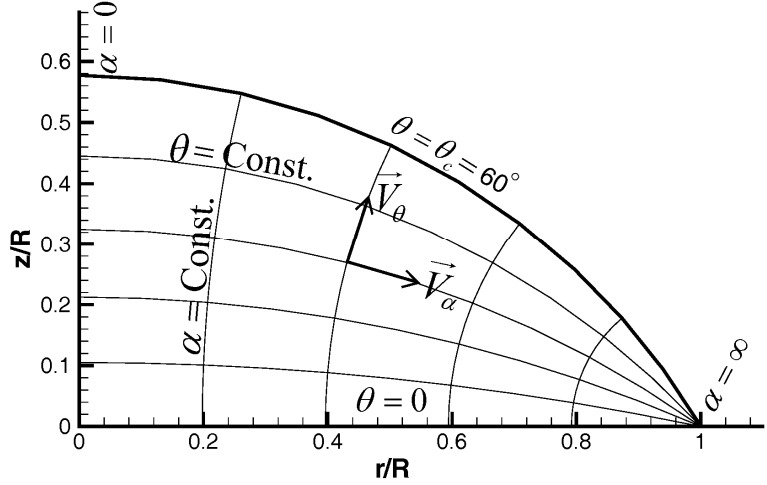

FIG. 1. Lines of constant $\alpha$ and $\theta$ and positive direction of velocity vector components in toroidal coordinates.

## B. Field equation

The flow within the liquid drop is incompressible. Hence:

$$\nabla \cdot \mathbf{V} = 0, \tag{3}$$

where $\mathbf{V}$ is the velocity vector. In addition, fluid is taken to be inviscid and the evaporation rate is assumed to be slow enough that the flow field in the droplet may be treated as quasi-steady. Consequently, the flow contains no vorticity:

$$\boldsymbol{\omega} = \nabla \times \mathbf{V} = 0. \tag{4}$$

A stream function, $\psi$, may be defined for this axially-symmetric flow such that it exactly satisfies Eq. (3):

$$V_\alpha = \frac{1}{h_\theta h_\varphi} \frac{\partial \psi}{\partial \theta} = \frac{(\cosh\alpha + \cos\theta)^2}{R^2 \sinh\alpha} \frac{\partial \psi}{\partial \theta}, \tag{5}$$

$$V_\theta = -\frac{1}{h_\varphi h_\alpha}\frac{\partial \psi}{\partial \alpha} = -\frac{(\cosh\alpha+\cos\theta)^2}{R^2 \sinh\alpha}\frac{\partial \psi}{\partial \alpha}. \qquad (6)$$

The field equation for $\psi$ follows from writing Eq. (4) for axially-symmetric flow and then substituting $V_\alpha$ and $V_\theta$ in terms of $\psi$ from Eqs. (5) and (6). In general, for axially-symmetric flows the vorticity is related to the stream function by:

$$\boldsymbol{\omega} = \frac{\hat{\mathbf{e}}_\varphi}{h_\alpha h_\theta}\left[\frac{\partial}{\partial \alpha}\left(\frac{h_\theta}{h_\varphi h_\alpha}\frac{\partial \psi}{\partial \alpha}\right) + \frac{\partial}{\partial \theta}\left(\frac{h_\alpha}{h_\theta h_\varphi}\frac{\partial \psi}{\partial \theta}\right)\right]. \qquad (7)$$

Inserting the metric coefficients for toroidal geometry from Eq. (1):

$$\boldsymbol{\omega} = \frac{\hat{\mathbf{e}}_\varphi}{R^3}(\cosh\alpha+\cos\theta)^2$$

$$\times \left[\frac{\partial}{\partial \alpha}\left(\frac{\cosh\alpha+\cos\theta}{\sinh\alpha}\frac{\partial \psi}{\partial \alpha}\right) + \frac{\partial}{\partial \theta}\left(\frac{\cosh\alpha+\cos\theta}{\sinh\alpha}\frac{\partial \psi}{\partial \theta}\right)\right],$$

$$\equiv (\hat{\mathbf{e}}_\varphi / h_\varphi)\, E^2\psi . \qquad (8)$$

For inviscid flow, the field equation for $\psi$ corresponds to $\boldsymbol{\omega}=0$. Or, as seen from the above equation, $E^2\psi = 0$.

### C. Integration of $E^2\psi = 0$

The discussion below presents the closed-form solution obtained by integrating $E^2\psi = 0$, subject to appropriate boundary conditions.

#### 1. Separation of variables

The $\alpha$ and $\theta$ variables may be separated in the field equation $E^2\psi = 0$ by letting

$$\psi(\alpha,\theta) = (\cosh\alpha+\cos\theta)^{-1/2}\, f(\alpha)\, g(\theta).$$

The sign of the resulting separation constant is chosen to obtain eigenfunctions in $\alpha$. Since $0 \leq \alpha < \infty$, the associated eigenvalues are continuously distributed: $0 \leq \tau < \infty$. As reported in [26], the solution is:

$$\psi(\alpha,\theta) = (\cosh\alpha + \cos\theta)^{-1/2} \int_0^\infty [a_1(\tau) C_{1/2+i\tau}^{-1/2}(\cosh\alpha) + a_2(\tau) C_{1/2+i\tau}^{*-1/2}(\cosh\alpha)]$$

$$\times [c_1(\tau)\sinh(\tau\theta) + c_2(\tau)\cosh(\tau\theta)]\, d\tau. \tag{9}$$

where, $C_{1/2+i\tau}^{-1/2}(x)$ and $C_{1/2+i\tau}^{*-1/2}(x)$ are Gegenbauer functions of the first and second kind, of order $-1/2$ [27]. Regarding the Gegenbauer function of the first kind, its properties relevant to the present study are given in the Appendix along the development of its integral transform and inverse which are needed.

## 2. Boundary conditions

To establish boundary conditions on the stream function, $\psi$, it is first noted that the velocity components normal to the axis of symmetry and normal to the solid surface vanish: $V_\alpha(0,\theta) = 0$ and $V_\theta(\alpha,0) = 0$. Hence, the stream function is constant along the symmetry axis and the solid surface. Since these lines intersect, the constant must be the same for both. The value of this constant does not affect the predicted velocities. Hence, may arbitrarily be set to zero. The corresponding boundary conditions are then:

$$\psi(\alpha,0) = 0 \tag{10}$$

and

$$\psi(0,\theta) = 0. \tag{11}$$

Equation (10) requires $c_2(\tau) = 0$. For Eq. (11), it is noted that $C_{1/2+i\tau}^{*-1/2}(\cosh\alpha)$ becomes infinite along the axis of symmetry and, therefore, it is required that $a_2(\tau) = 0$.

Therefore, the form of the solution is:

$$\psi(\alpha,\theta) = (\cosh\alpha + \cos\theta)^{-1/2} \int_0^\infty k(\tau)\, C_{1/2+i\tau}^{-1/2}(\cosh\alpha)\, \sinh(\tau\theta)\, d\tau. \tag{12}$$

The second $\alpha$–boundary condition is:

$$\psi(\infty,\theta) = \text{finite}. \tag{13}$$

which is automatically satisfied.

The unknown coefficient $k(\tau)$ can be written in terms of the stream function at the free surface using an infinite integral transform. This transform is based on the Gegenbauer function as the eigenfunction. It is similar to the Mehler-Fock transform [28] which is based on Legendre functions. The required Gegenbauer transform could not be found in the literature and so they were derived as part of the present study. Derivation of the required transform and its inverse is presented in the Appendix. In particular, the integral relation between the unknown coefficient $k(\tau)$ and the stream function at the free surface is obtained from using Eqs. (A7) and (A8) in the Appendix. The result is:

$$k(\tau) = \frac{\tau(\tau^2 + 1/4)\tanh(\pi\tau)}{\sinh(\tau\theta_c)}$$

$$\times \int_0^\infty \frac{\psi(\alpha,\theta_c)\,(\cosh\alpha + \cos\theta_c)^{1/2}}{\sinh\alpha}\, C_{1/2+i\tau}^{-1/2}(\cosh\alpha)\, d\alpha. \tag{14}$$

The final boundary condition must therefore relate the stream function at the free surface to a specified distribution of evaporative mass flux. This flux is determined by analyzing the gas phase transport of the evaporating species. At the liquid/gas interface, the mass flux of the evaporating species is the same in each phase. In terms of the liquid, this flux may be written as:

$$J(\alpha) = \rho\,[V_\theta(\alpha,\theta_c) - V_{\theta,B}(\alpha)], \tag{15}$$

where, $J(\alpha)$ is the evaporative flux at the free surface, $\rho$ is the liquid density and $V_{\theta,B}$ is the speed at which the boundary is moving in the direction normal to itself. From Eq. (6), the boundary condition on $\psi$ may then be written in terms of $V_\theta(\alpha,\theta)\big|_{\theta=\theta_c}$:

$$\psi(\alpha,\theta_c) = -\int_0^\alpha \frac{R^2 \sinh\alpha'}{(\cosh\alpha' + \cos\theta_c)^2} V_\theta(\alpha',\theta_c)\, d\alpha'$$

$$= -\int_0^\alpha \frac{R^2 \sinh\alpha'}{(\cosh\alpha' + \cos\theta_c)^2} [J(\alpha')/\rho + V_{\theta,B}(\alpha')]\, d\alpha'. \quad (16)$$

## III. PINNED CONTACT LINE

When the droplet is 'pinned' at the contact line, $R$ is constant and $\theta_c = \theta_c(t)$. Then the differential length through which a $d\alpha$ element of free surface moves during evaporation is given by $dx_B(\alpha) = [h_\theta(\alpha,\theta)\, d\theta]\big|_{\theta_c}$. The corresponding speed of the boundary, $V_B = dx_B/dt$, is therefore

$$V_{\theta,B}(\alpha) = R\,(\cosh\alpha + \cos\theta_c)^{-1}(d\theta_c/dt). \quad (17)$$

In terms of an arbitrary evaporative flux distribution, the interface stream function may be written from Eqs. (16) and (17) as:

$$\psi(\alpha,\theta_c) = -\frac{d\theta_c}{dt}\int_0^\alpha \frac{R^3 \sinh\alpha'}{(\cosh\alpha' + \cos\theta_c)^3}\, d\alpha'$$

$$-\int_0^\alpha \frac{R^2 \sinh\alpha'}{(\cosh\alpha' + \cos\theta_c)^2} \frac{J(\alpha')}{\rho}\, d\alpha', \quad (18)$$

Upon completing the first integral, this reduces to

$$\psi(\alpha,\theta_c) = -\frac{d\theta_c}{dt}\frac{R^3}{2}\left[\frac{1}{(1+\cos\theta_c)^2} - \frac{1}{(\cosh\alpha+\cos\theta_c)^2}\right]$$

$$-\int_0^\alpha \frac{R^2 \sinh\alpha'}{(\cosh\alpha' + \cos\theta_c)^2} \frac{J(\alpha')}{\rho} d\alpha' \qquad (18')$$

For the stream function to be continuous between the free surface and the substrate, the stream function must be zero as the contact line is approached along the free surface; that is, $\psi(\alpha \to \infty; \theta = \theta_c) = 0$. Using this behavior in the above equation and rearranging yields:

$$\frac{d\theta_c}{dt} = -(2/R)(1+\cos\theta_c)^2 \int_0^\infty \frac{\sinh\alpha}{(\cosh\alpha + \cos\theta_c)^2} \frac{J(\alpha)}{\rho} d\alpha \qquad (19)$$

A wide range of evaporative flux distributions is possible depending on the velocity distributions in the gas phase and the degree of vacuum into which the evaporation occurs. In the following sections two specific cases are investigated.

### A. Uniform evaporative flux

In this case, the evaporation rate is uniform across the surface of the drop:

$$J(\alpha, \theta_c) = \text{const.} = J_0. \qquad (20)$$

Consequently, Eq. (19) reduces to

$$\frac{d\theta_c}{dt} = -\frac{2J_0}{R\rho}(1+\cos\theta_c) \qquad (21)$$

Using Eqs. (20) and (21), Eq. (18') can be written as

$$\psi(\alpha, \theta_c) = \frac{R^2 J_0}{\rho\,(\cosh\alpha + \cos\theta_c)}\left(1 - \frac{1+\cos\theta_c}{\cosh\alpha + \cos\theta_c}\right) \qquad (22)$$

and, using Eq. (14) in conjunction with Eq. (A15) of the Appendix, the expansion coefficient becomes

$$k(\tau) = (R^2 J_0 / \rho)\sqrt{2}\ \tau\ \mathrm{cosec}\,\theta_c\ \mathrm{sech}(\pi\tau)$$

$$\times \{1 - 2(1+\cos\theta_c)\mathrm{cosec}\,\theta_c\,[\tau\coth(\tau\theta_c) - \cot\theta_c]\} \tag{23}$$

## B. Diffusive evaporative flux

A commonly considered flux distribution corresponds to diffusive mass transfer into a stagnant gas (*i.e.*, absent even the gas motion which naturally occurs due to the mass transfer). This model leads to Laplace's equation in toroidal coordinates for the variation of vapor concentration throughout the gas phase. Solving this equation results in the following evaporative mass flux distribution (see Popov [13]):

$$J(\alpha,\theta_c) = (Y_s - Y_\infty)(\rho_g D / R)\Big\{\sin\theta_c / 2 + \sqrt{2}(\cosh\alpha + \cos\theta_c)^{3/2}$$

$$\times \int_0^\infty \frac{\cosh\theta_c\tau}{\cosh\pi\tau} \tanh[(\pi-\theta_c)\tau]\,P_{-1/2+i\tau}(\cosh\alpha)\,\tau\,d\tau\Big\}, \tag{24}$$

where $P_{-1/2+i\tau}(x)$ is the conical function of the first kind, $\rho_g$ is the density of the gaseous vapor-air mixture, $D$ is the coefficient of binary diffusion of the vapor in the gas phase, $Y_s$ is the vapor mass fraction in the gas phase at the droplet surface (saturation value) and $Y_\infty$ is the far field vapor mass fraction in the gas phase.

Using Eq. (24) and Eq. (7-6-9) of Ref. [28], Eq. (19) reduces to

$$\frac{d\theta_c}{dt} = -\frac{\rho_g D(Y_s - Y_\infty)}{\rho R^2}(1+\cos\theta_c)^2$$

$$\times \left\{\frac{\sin\theta_c}{1+\cos\theta_c} + 4\int_0^\infty \frac{1+\cosh 2\theta_c\tau}{\sinh 2\pi\tau}\tanh[(\pi-\theta_c)\tau]\,d\tau\right\}, \tag{25}$$

(Note that Popov [13] used a different approach for determining $d\theta_c / dt$.)

In the next section, particular cases of the above general results are considered. Specifically, hemispherical drops and drops having small contact angles ($\theta_c \simeq 0$) are treated with the following being given in each case: the evaporation rate, $J(\alpha)$, the rate of change of contact angle, ($d\theta_c / dt$), for 'pinned' contacts, the distribution of the stream function at the free surface, $\psi(\alpha,\theta_c)$, and the expansion coefficient, $k(\tau)$.

### 1. Hemishperical shape, $\theta_c = \pi/2$

For the hemispherical shape, $\theta_c = \pi/2$, the diffusive evaporative flux is uniformly distributed over the surface of the drop [13,21]:

$$J(\alpha) = \frac{\rho_g D (Y_s - Y_\infty)}{R}. \tag{26}$$

Eq. (25) therefore becomes:

$$\frac{d\theta_c}{dt} = -\frac{\rho_g D (Y_s - Y_\infty)}{\rho (R^2 / 2)}, \tag{27}$$

As a result, the stream function distribution along the free surface is given by:

$$\psi(\alpha,\pi/2) = \frac{\rho_g R D (Y_s - Y_\infty)}{\rho \cosh \alpha}\left(1 - \frac{1}{\cosh \alpha}\right), \tag{28}$$

Using Eq. (14) in conjunction with Eq. (A15) of the Appendix, the spectral coefficient is given by:

$$k(\tau) = (\rho_g / \rho) R D (Y_s - Y_\infty)\sqrt{2}\, \tau\, \text{sech}(\pi \tau)[1 - 2\tau \coth(\tau \pi / 2)]. \tag{29}$$

### 2. Small contact angle, $\theta_c \to 0$

For contact angles small enough such that $\cos\theta_c \simeq 1$ and $\sin\theta_c \simeq \theta_c \simeq 0$, it is known that [13]

$$J(\alpha) = \frac{\rho_g D(Y_s - Y_\infty)}{R} \frac{\sqrt{2}}{\pi} \sqrt{\cosh\alpha + 1}, \qquad (30)$$

which reduces Eq. (25) to:

$$\frac{d\theta_c}{dt} = -\frac{16\rho_g D(Y_s - Y_\infty)}{\pi \rho R^2} \qquad (31)$$

and, consequently,

$$\psi(\alpha,\theta_c) = \frac{2\rho_g R D(Y_s - Y_\infty)}{\pi \rho}\left[\sqrt{\frac{2}{\cosh\alpha+1}} - \frac{4}{(\cosh\alpha+1)^2}\right] \qquad (32)$$

Using Eq. (14) in conjunction with Eq. (A15) of the Appendix, the spectral coefficient is then given by:

$$k(\tau) = -\frac{16\sqrt{2}\rho_g R D(Y_s - Y_\infty)}{3\pi \rho} \frac{\tau^2(\tau^2+1)}{\cosh(\pi\tau)\sinh(\theta_c \tau)} \qquad (33)$$

## IV. FREELY MOVING CONTACT LINE

When the drop is 'freely moving', the radial distance to the contact line decreases with time $R = R(t)$. One condition previously considered for the freely moving case is that the contact angle remains constant during evaporation ($\theta_c = \text{constant}$). The speed of a spherical cap boundary in the normal direction is the same as for the cylindrical cap boundary analyzed by Petsi and Burganos [23] since both have the same cross sectional shape. This speed is given by:

$$V_{\theta,B}(\alpha) = \sin\theta_c \cosh\alpha(\cosh\alpha + \cos\theta_c)^{-1}(dR/dt). \qquad (34)$$

In terms of an arbitrary evaporative flux distribution, the interface stream function may be written from Eqs. (16) and (34) as:

$$\psi(\alpha,\theta_c) = -\frac{dR}{dt}\int_0^\alpha \frac{R^2 \sin\theta_c \cosh\alpha' \sinh\alpha'}{(\cosh\alpha' + \cos\theta_c)^3} d\alpha'$$

$$-\int_0^\alpha \frac{R^2 \sinh\alpha'}{(\cosh\alpha' + \cos\theta_c)^2} \frac{J(\alpha')}{\rho} d\alpha', \tag{35}$$

which, upon completing the first integral, reduces to

$$\psi(\alpha,\theta_c) = -\frac{dR}{dt} R^2 \sin\theta_c \left[\frac{1+\cos\theta_c/2}{(1+\cos\theta_c)^2} - \frac{\cosh\alpha + \cos\theta_c/2}{(\cosh\alpha + \cos\theta_c)^2}\right]$$

$$-\int_0^\alpha \frac{R^2 \sinh\alpha'}{(\cosh\alpha' + \cos\theta_c)^2} \frac{J(\alpha')}{\rho} d\alpha'. \tag{35'}$$

For the stream function to be continuous, its value on the free surface must vanish as the contact line is approached; that is, $\psi(\alpha \to \infty; \theta = \theta_c) = 0$. Using this in the above equation and rearranging yields:

$$\frac{dR}{dt} = -\frac{(1+\cos\theta_c)^2}{\sin\theta_c(1+\cos\theta_c/2)} \int_0^\infty \frac{\sinh\alpha}{(\cosh\alpha + \cos\theta_c)^2} \frac{J(\alpha)}{\rho} d\alpha. \tag{36}$$

Of the wide range of evaporative flux distributions is possible, the following sections investigate two specific cases: uniform flux and the flux corresponding to purely diffusive gas phase mass transfer.

## A. Uniform evaporative flux

In this case, the evaporation rate is uniform over the surface of the drop, that is,

$$J(\alpha,\theta_c) = \text{const.} = J_0. \tag{37}$$

Consequently, Eq. (36) reduces to:

$$\frac{dR}{dt} = -\frac{J_0}{\rho} \frac{1+\cos\theta_c}{\sin\theta_c(1+\cos\theta_c/2)}. \tag{38}$$

Using Eqs. (37) and (38), Eq. (35′) can be written as

$$\psi(\alpha,\theta_c) = -\frac{R^2 J_0}{\rho} \frac{\cos\theta_c}{(\cosh\alpha+\cos\theta_c)(2+\cos\theta_c)}\left(1-\frac{1+\cos\theta_c}{\cosh\alpha+\cos\theta_c}\right). \tag{39}$$

Then, using Eq. (14) in conjunction with Eq. (A15) of the Appendix, the expansion coefficient becomes:

$$k(\tau) = -(R^2 J_0/\rho)\sqrt{2}\,\tau\,\operatorname{cosec}\theta_c\,\operatorname{sech}(\pi\tau)\cos\theta_c(2+\cos\theta_c)^{-1}$$

$$\times\left\{1 - 2(1+\cos\theta_c)\operatorname{cosec}\theta_c[\tau\coth(\tau\theta_c) - \cot\theta_c]\right\} \tag{40}$$

### B. Diffusive evaporative flux

The evaporative flux distribution is again given by Eq. (24), which is repeated here:

$$J(\alpha,\theta_c) = (Y_s - Y_\infty)(\rho_g D/R)\Bigg\{\sin\theta_c/2 + \sqrt{2}(\cosh\alpha+\cos\theta_c)^{3/2}$$

$$\times \int_0^\infty \frac{\cosh\theta_c\tau}{\cosh\pi\tau}\tanh[(\pi-\theta_c)\tau]\,P_{-1/2+i\tau}(\cosh\alpha)\,\tau\,d\tau\Bigg\},$$

Using Eq. (24) and Eq. (7-6-9) of Ref. [28], Eq. (36) reduces to

$$\frac{dR}{dt} = -\frac{\rho_g D(Y_s - Y_\infty)}{\rho R}\frac{(1+\cos\theta_c)^2}{\sin\theta_c(2+\cos\theta_c)}$$

$$\times\left\{\frac{\sin\theta_c}{1+\cos\theta_c} + 4\int_0^\infty \frac{1+\cosh 2\theta_c\tau}{\sinh 2\pi\tau}\tanh[(\pi-\theta_c)\tau]d\tau\right\}. \tag{41}$$

Given below for freely moving contact lines in the limits of hemispherical shape and small contact angle ($\theta_c \simeq 0$) are the following: the evaporation rate, $J(\alpha)$, the rate of change of droplet radius, ($dR/dt$), the stream function on the free surface, $\psi(\alpha,\theta_c)$, and the expansion coefficient, $k(\tau)$.

### 1. Hemispherical shape, $\theta_c = \pi/2$

Being hemispherical, the evaporative flux distribution is uniform and is given again by Eq. (26):

$$J(\alpha) = \frac{\rho_g D (Y_s - Y_\infty)}{R}.$$

Eq. (41) then becomes:

$$\frac{dR}{dt} = -\frac{\rho_g D (Y_s - Y_\infty)}{\rho R}. \tag{42}$$

Consequently, the stream function distribution along the free surface becomes:

$$\psi(\alpha, \pi/2) = 0. \tag{43}$$

Therefore, the spectral coefficient for the freely moving condition is:

$$k(\tau) = 0 \tag{44}$$

### 2. Small contact angle, $\theta_c \to 0$

For contact angles small enough such that $\cos\theta_c \simeq 1$ and $\sin\theta_c \simeq \theta_c \simeq 0$, it is known that [13]

$$J(\alpha) = \frac{\rho_g D(Y_s - Y_\infty)}{R} \frac{\sqrt{2}}{\pi} \sqrt{\cosh\alpha + 1},$$

which reduces Eq. (41) to

$$\frac{dR}{dt} = -\frac{16 \rho_g D(Y_s - Y_\infty)}{3\pi \theta_c \rho R}. \tag{45}$$

and, consequently,

$$\psi(\alpha,\theta_c) = \frac{2\rho_g R D(Y_s - Y_\infty)}{\pi \rho}\left[\sqrt{\frac{2}{\cosh\alpha+1}} - \frac{4(2\cosh\alpha+1)}{3(\cosh\alpha+1)^2}\right]. \tag{46}$$

Using Eq. (14) in conjunction with Eq. (A15) of the Appendix, the spectral coefficient is given by:

$$k(\tau) = \frac{16\sqrt{2}\rho_g R D(Y_s - Y_\infty)}{3\pi\rho} \frac{\tau^2[(\tau^2+1)/3 - 1]}{\cosh(\pi\tau)\sinh(\theta_c\tau)}. \tag{47}$$

## V. VELOCITY DISTRIBUTION

Given the distribution of evaporative flux, $J(\alpha)$, the boundary stream function, $\psi(\alpha,\theta_c)$, may be determined from Eq. (16) for given distribution of the velocity normal to the boundary, $V_{\theta,B}$ (e.g., corresponding to contact lines which are either pinned or freely moving, Eqs. (17) and (34), respectively). Then, the coefficient $k(\tau)$ in the eigenfunction expansion for the stream function $\psi(\alpha,\theta)$ may be evaluated from Eq. (14). Finally, from $\psi(\alpha,\theta)$, the velocity distribution may be calculated. In toroidal coordinates the components of the velocity follow from Eqs. (5) and (6):

$$V_\alpha(\alpha,\theta) = \frac{(\cosh\alpha+\cos\theta)^{3/2}}{R^2 \sinh\alpha}\left\{\frac{\sin\theta}{2}\frac{\psi(\alpha,\theta)}{(\cosh\alpha+\cos\theta)^{1/2}}\right.$$

$$\left. + \int_0^\infty k(\tau)\tau\cosh(\tau\theta)C_{1/2+i\tau}^{-1/2}(\cosh\alpha)d\tau\right\}, \tag{48}$$

$$V_\theta(\alpha,\theta) = \frac{(\cosh\alpha+\cos\theta)^{3/2}}{R^2}\left\{\frac{\psi(\alpha,\theta)}{2(\cosh\alpha+\cos\theta)^{1/2}}\right.$$

$$+\int_0^\infty k(\tau)\sinh(\tau\theta)P_{-1/2+i\tau}(\cosh\alpha)d\tau\Bigg\}. \tag{49}$$

The velocity components in cylindrical coordinates are useful for visualizing and physically interpreting the flow field. The radial and axial components of the velocity are given by:

$$V_r(\alpha,\theta) = r^{-1}(\partial\psi/\partial z)$$
$$= (\cosh\alpha+\cos\theta)^{-1}\left[V_\alpha(1+\cosh\alpha\cos\theta)+V_\theta\sinh\alpha\sin\theta\right], \tag{50}$$

$$V_z(\alpha,\theta) = -r^{-1}(\partial\psi/\partial r)$$
$$= -(\cosh\alpha+\cos\theta)^{-1}\left[V_\alpha\sinh\alpha\sin\theta-V_\theta(1+\cosh\alpha\cos\theta)\right]. \tag{51}$$

## VI.  RESULTS AND DISCUSSION

The flow patterns corresponding to pinned and freely moving contact lines are computed for both uniform and diffusive evaporative flux distributions. Comparisons are made for the cases of wetting, non-wetting and hemispherical drops. The computed two-dimensional velocities are compared to the radial distributions of the vertically averaged velocities used in previous analyses [13,15-17],

$$<V_r>(r) = \frac{1}{h(r)}\int_0^{h(r)} V_r(r,z)\,dz, \tag{52}$$

where $V_r$ is the radial component of the velocity and $h(r)$ is the thickness of the drop at a distance $r$ from the axis of symmetry.

The results are presented in terms of the dimensionless stream function, $\psi^* = \psi / \psi_0$, dimensionless velocity, $V^* = V / V_0$, and dimensionless evaporative flux, $J^* = J / J_0$, in which,

$$\psi_0 = R^2 J_0 / \rho, \tag{53}$$

$$V_0 = J_0 / \rho, \tag{54}$$

and $J_0$ is the characteristic evaporative flux where, for diffusive evaporation it is determined as

$$J_0 = \rho_g D (Y_s - Y_\infty) / R. \tag{55}$$

## A. Diffusive evaporative flux

The evaporation rate corresponding to diffusion through a stagnant gas is shown in Fig. 2(a) for contact angles 120°, 90° and 60°. These contact angles have been chosen to span the range from non-wetting to wetting behavior. For contact angles greater than ninety degrees, the evaporation rate decays to zero as $(r/R) \to 1$. For $\theta_c = \pi/2$, the evaporation rate is uniform over the free surface. For contact angles less than ninety degrees, the evaporation rate diverges at the contact line.

The flow generated by the evaporation is illustrated for contact angles 120°, 90° and 60°. Pinned and freely moving contact line behaviors are shown in Figs. 2(b), 3 and 4. When the contact line is pinned, the flow is directed from the center of the drop to its edges (for colloidal suspensions, this produces 'coffee-ring'–like deposits). The character of 'pinned' flow remains the same even for contact angles greater than ninety degrees where the evaporative flux distribution is quite different. As expected, flow field calculations for

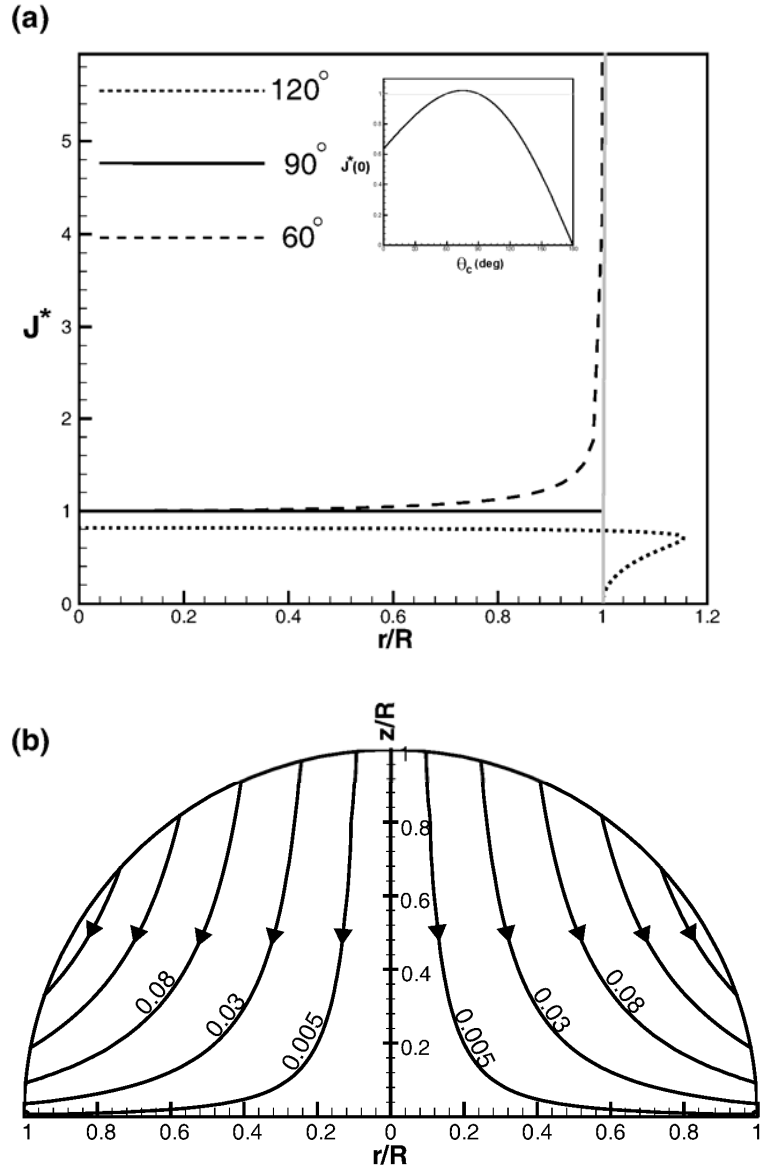

FIG 2. (a) Nondimensional diffusive evaporative flux for contact angles of 120°, 90° and 60°; Inset: flux at $r/R = 0$ versus contact angle (b) Hemispherical ($\theta_c = 90°$); contours of nondimensional stream function ($\psi^* = 0.005, 0.03, 0.08, 0.15$ and $0.22$) for pinned contact and uniform diffusive evaporative flux.

$\theta_c \to \pi/2$ coincide with those obtained from analyses performed in spherical coordinates [21]. It is also noteworthy that for $\theta_c = \pi/2$ and freely moving contact lines, the velocity inside the drop vanishes. This occurs because $V_{\theta,B}(\alpha) = -J(\alpha)/\rho$ for spherical drops. On the other hand, when the contact line is freely moving the flow pattern is more complicated. Flow both toward and away from the edge exist within the drop. In these cases it seems unlikely that 'coffee ring' type deposition of particles would occur during evaporation of a colloidal drop.

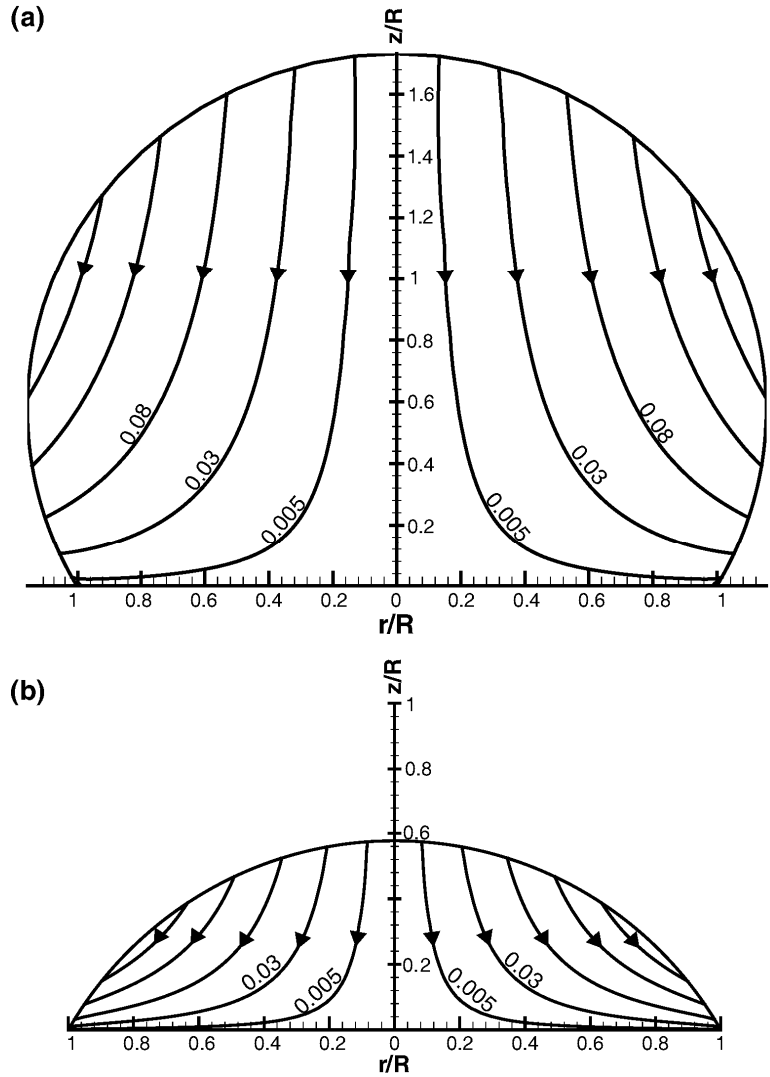

FIG. 3. Contours of nondimensional stream function ($\psi^* = 0.005, 0.03, 0.08, 0.15$ and $0.22$) for a pinned contact line, diffusive evaporative flux and contact angles of (a) 120° and (b) 60°.

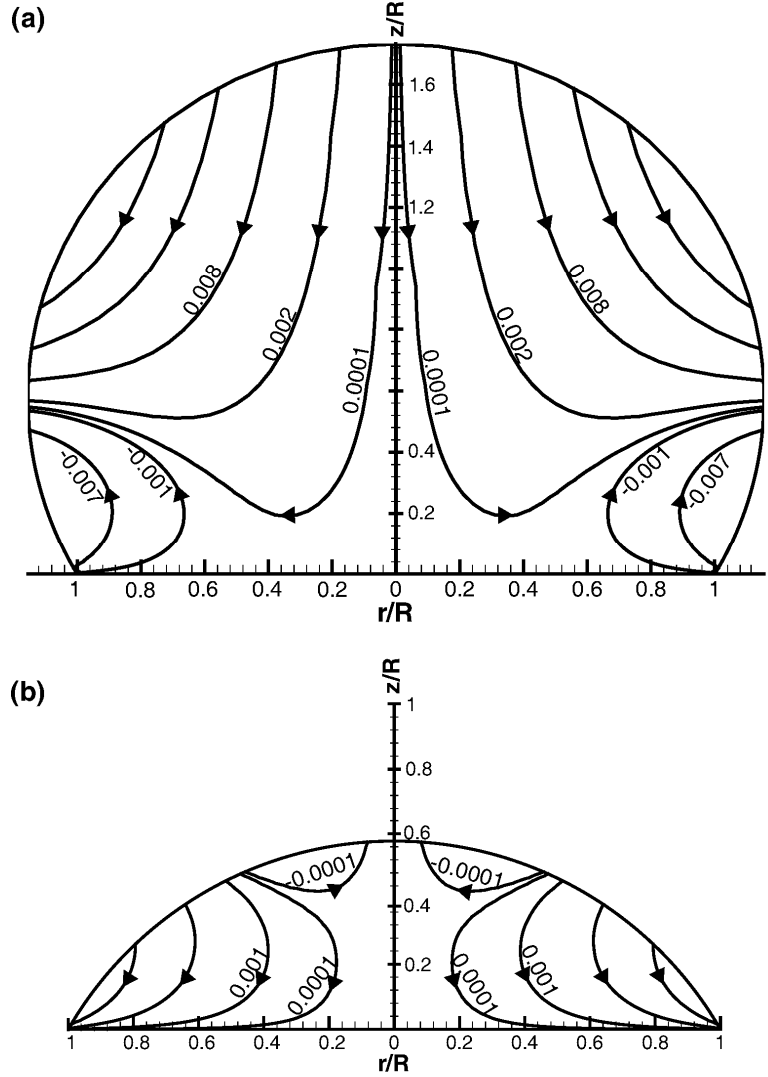

FIG. 4. Contours of nondimensional stream function for an freely moving contact line, diffusive evaporative flux and contact angles of (a) 120° ($\psi^* = $ -0.007, -0.001, 0.0001, 0.002, 0.008, 0.017 and 0.027) and (b) 60° ($\psi^* = $ -0.0001, 0.0001, 0.001, 0.006 and 0.02).

## B. Uniform evaporative flux

For uniform evaporative flux, the flow patterns computed for different contact angles and contact line conditions are illustrated in Figs. 5 and 6. Figure 5 shows that when the contact line is pinned, the flow is from the center of the drop to its edge. On the other hand, when the contact line is free to move, distinctively different flow patterns are observed for wetting ($\theta_c < \pi/2$) and non-wetting ($\theta_c > \pi/2$) conditions – see Fig. 6. (This is consistent with the observation previously made for cylindrical caps [23].) For contact angles greater than $\pi/2$, the flow pattern for an freely moving contact line is similar to a pinned contact line – Fig. 5(a). However, for contact angles less than $\pi/2$, the flow in the freely moving case is from the edge of the drop towards its axis – opposite the flow behavior for the pinned case.

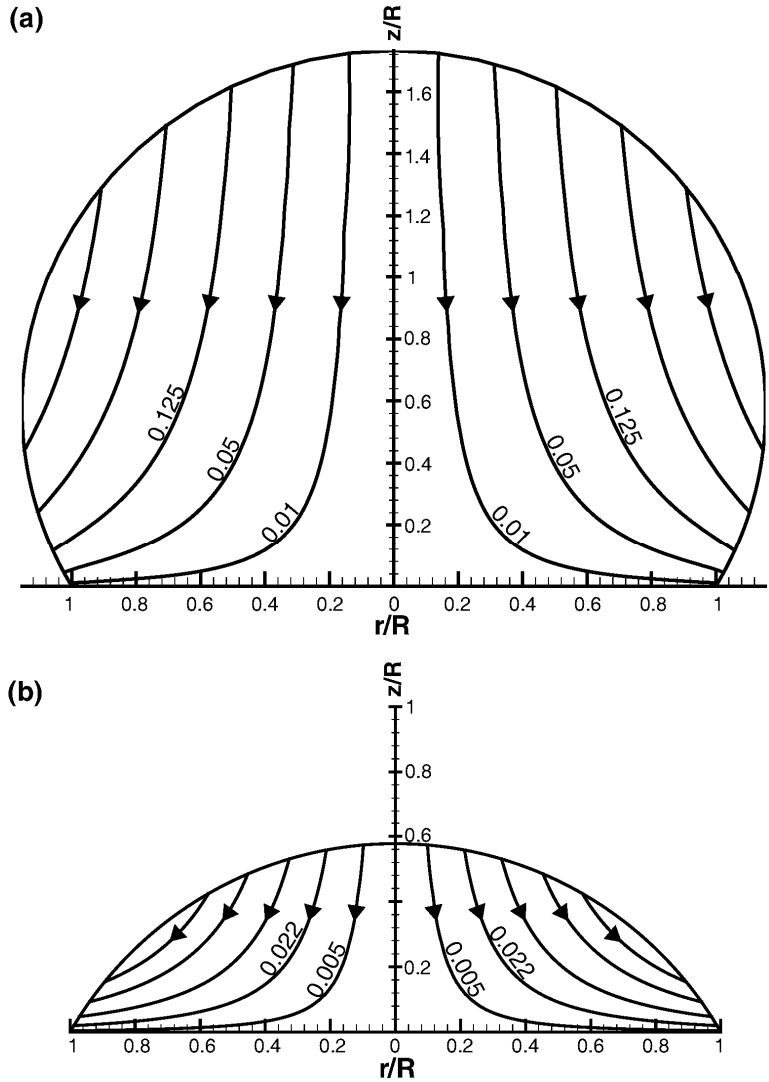

FIG. 5. Contours of nondimensional stream function for a pinned contact line, uniform evaporative flux and contact angles of (a) 120° ($\psi^* = 0.01$, 0.05, 0.125, 0.24 and 0.38) and (b) 60° ($\psi^* = 0.005$, 0.022, 0.05, 0.09 and 0.13).

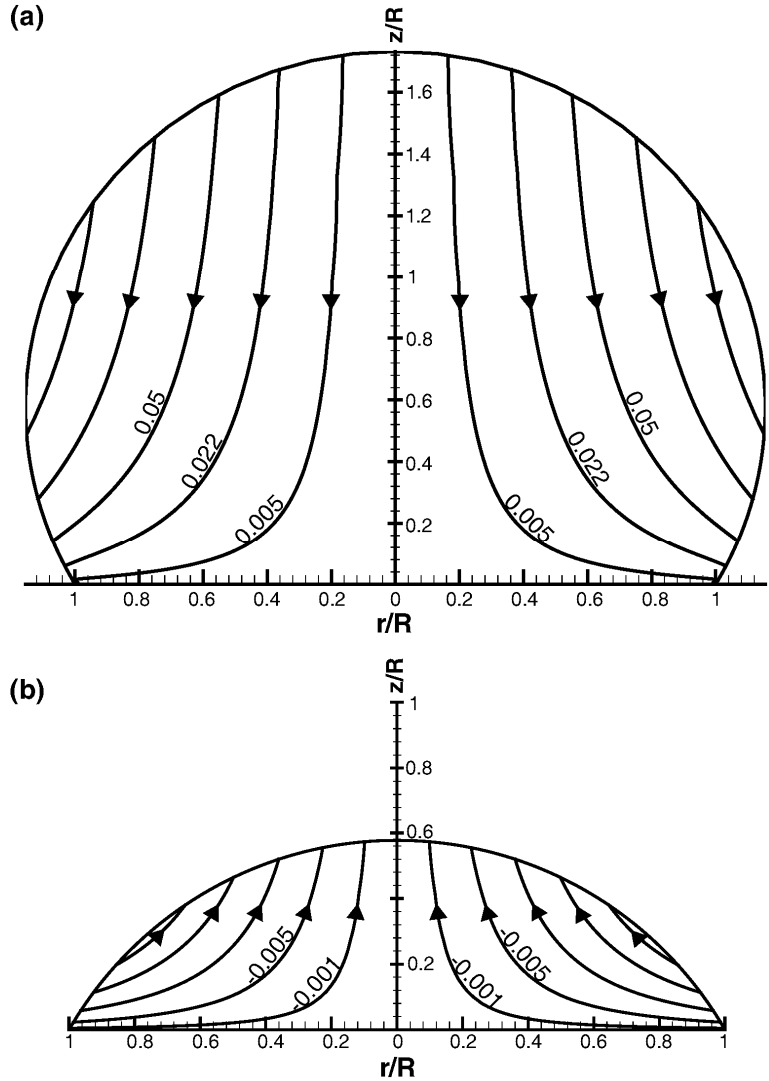

FIG. 6. Contours of nondimensional stream function for an freely moving contact line, uniform evaporative flux and contact angles of (a) 120° ($\psi^* = 0.005, 0.022, 0.05, 0.09$ and $0.135$) and (b) 60° ($\psi^* = -0.001, -0.005, -0.012, -0.021$ and $-0.03$).

Finally, it is to be noted that when the drop wets the surface, the flow is directed from the center to the edge for a pinned contact line, and from the edge towards the center when the contact line is free to move. On the other hand, for non-wetting drops, the flow is directed towards the edge for both pinned and freely moving contact lines.

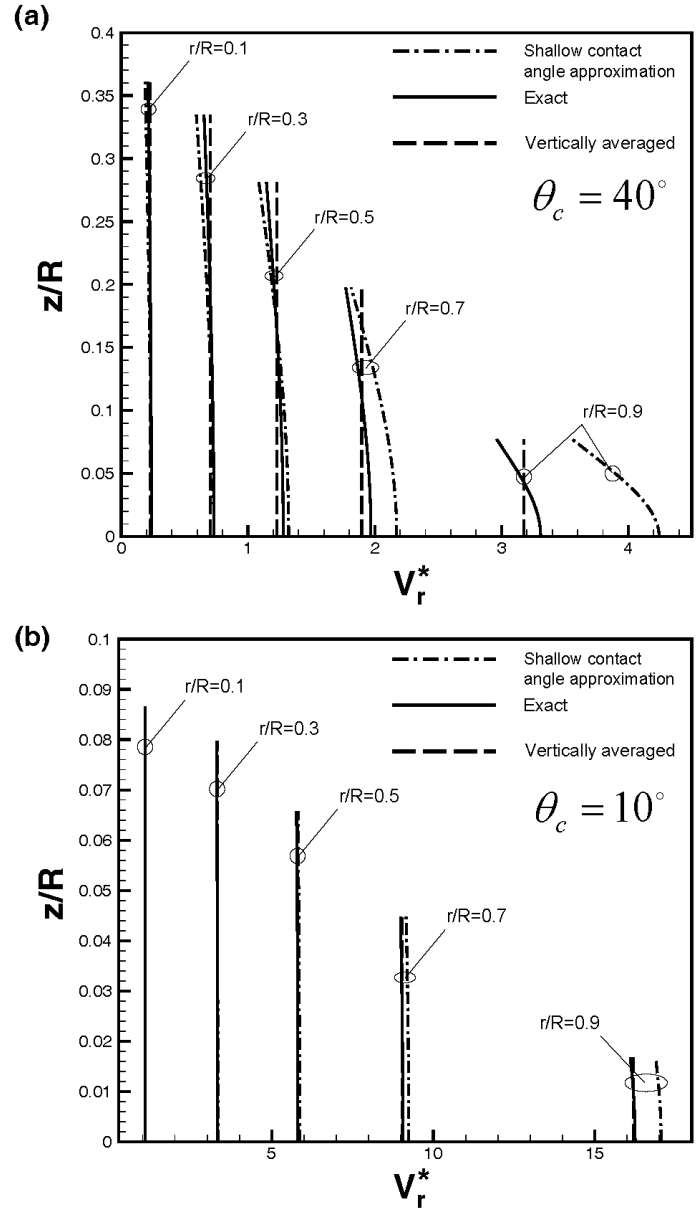

FIG . 7. Nondimensional radial velocities versus vertical position at different radial positions r/R= 0.1, 0.3, 0.5, 0.7 and 0.9, for a pinned contact line, diffusive evaporative flux and contact angles of (a) 40° and (b) 10°. The dash-dot lines are from the small contact angle approximation, the solid lines are from the exact solution, and the dashed lines are from the vertically averaged radial velocity.

### C. Approximate analyses compared to the exact result

At various radii, Fig. 7 compares the vertically averaged radial velocity, Eq. (52),

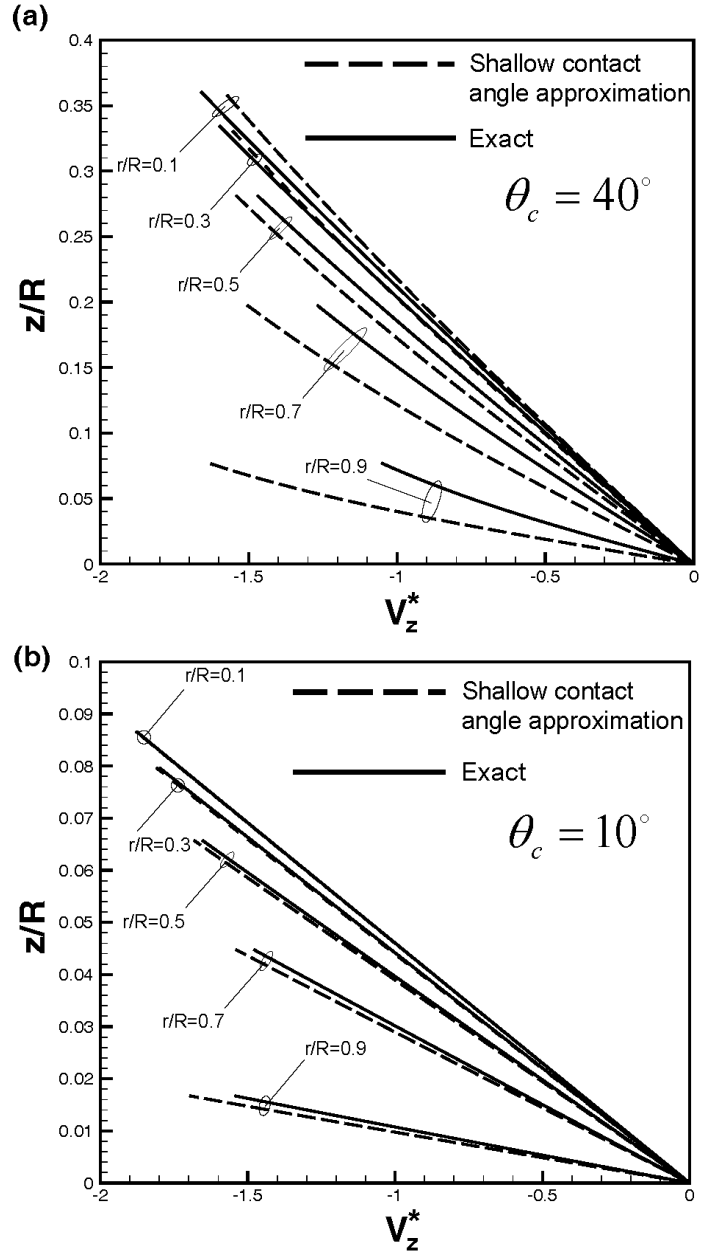

FIG 8. Nondimensional vertical velocities versus vertical position at different radial positions r/R= 0.1, 0.3, 0.5, 0.7 and 0.9, for a pinned contact line, diffusive evaporative flux and contact angles of (a) 40° and (b) 10°. The dashed lines are from the small contact angle approximation and the solid lines are from the exact solution.

to the z-variations of the exact solution and the small contact angle approximation. Shown are the results for pinned contacts with contact angles of 40° and 10°. It is seen that at higher contact angles, only the exact solution faithfully represents the flow, particularly at radii away from the centerline. However, at small contact angles, Fig. 7(b) indicates a sharp decrease in the vertical variation of the radial velocity. Consequently, the vertically averaged velocity approach becomes an excellent approximation of the entire flow field at relatively small contact angles. On the other hand, Figs. 7 and 8 demonstrate that the small contact angle approximation for pinned contacts, Eq. (33), is only a good approximation to the exact solution for modest values of $(r/R)$ and small contact angles.

## ACKNOWLEDGMENT

The importance of this problem was brought to our attention by Prof. R.C. Wetherhold.

## APPENDIX: GEGENBAUER FUNCTION AND ITS INTEGRAL TRANSFORM

Several mathematical relations involving Gegenbauer functions were needed but could not be found in the literature. This appendix develops those relations. Throughout out the Appendix, $x = \cosh\alpha$ (Correspondingly, $\alpha = \cosh^{-1} x$, $\sinh\alpha = (x^2 - 1)^{1/2}$, and $dx = \sinh\alpha\, d\alpha$).

The Gegenbauer functions satisfy the following singular Sturm-Liouville differential equation

$$(x^2 - 1) f''(x) + (\tau^2 + 1/4) f(x) = 0, \tag{A1}$$

where $\tau$ is a parameter (eigenvalue) and the weighting function for normalization is $(x^2 - 1)^{-1}$. The range of $x$ in the present problem is $[1, \infty)$. The orthogonality/normalization condition for the Gegenbauer function is therefore

$$N(\tau_1, \tau_2) = \int_1^\infty (x^2 - 1)^{-1} C_{1/2+i\tau_1}^{-1/2}(x) C_{1/2+i\tau_2}^{-1/2}(x) \, dx \ . \tag{A2}$$

This may be evaluated from the orthogonality/normalization condition of Legendre functions [29] (weighting function = 1):

$$\int_1^\infty P_{-1/2+i\tau_1}(x) P_{-1/2+i\tau_2}(x) \, dx = \frac{\delta(\tau_1 - \tau_2)}{\tau_1 \tanh(\pi \tau_1)}. \tag{A3}$$

Note that [27]:

$$P_{-1/2+i\tau}(x) = -\frac{\partial}{\partial x} C_{1/2+i\tau}^{-1/2}(x) \tag{A4}$$

and

$$(x^2 - 1) \frac{\partial}{\partial x} P_{-1/2+i\tau}(x) = (\tau^2 + 1/4) C_{1/2+i\tau}^{-1/2}(x) \ . \tag{A5}$$

First, replace $P_{-1/2+i\tau_1}(x)$ in Eq. (A3) by Eq. (A4). Then, integrate by parts followed by replacing $\frac{\partial}{\partial x} P_{-1/2+i\tau_2}(x)$ in the resulting integral by Eq. (A5). This yields the following expression for the orthogonality/normalization of the Gegenbauer functions:

$$\int_1^\infty (x^2 - 1)^{-1} C_{1/2+i\tau_1}^{-1/2}(x) C_{1/2+i\tau_2}^{-1/2}(x) \, dx = \frac{\delta(\tau_1 - \tau_2)}{\tau_1 \tanh(\pi \tau_1)(\tau_1^2 + 1/4)}. \tag{A6}$$

It then follows that the transform of a function, $f(x)$, defined by :

$$f^*(\tau) = \int_1^\infty f(x)(x^2 - 1)^{-1} C_{1/2+i\tau}^{-1/2}(x) \, dx, \tag{A7}$$

has its inverse transform given by:

$$f(x) = \int_0^\infty \tau(\tau^2 + 1/4)\tanh(\pi\tau) f^*(\tau) \, C_{1/2+i\tau}^{-1/2}(x) \, d\tau. \tag{A8}$$

In our experience, an efficient way for computing the Gegenbauer function is in terms of an integral relation. This relation follows from substituting the integral representation of the Legendre function (Eq. (7.4.2) of [25]):

$$P_\nu(x) = \frac{1}{\pi}\int_0^\pi [x + (x^2-1)^{1/2}\cos\gamma]^{-(\nu+1)} \, d\gamma. \tag{A9}$$

into the following relation between Gegenbauer and Legendre functions [27]:

$$C_{1/2+i\tau}^{-1/2}(x) = \frac{1}{2i\tau}[P_{-3/2+i\tau}(x) - P_{1/2+i\tau}(x)], \tag{A10}$$

This yields the following integral representation of the Gegenbauer function:

$$C_{1/2+i\tau}^{-1/2}(x) = \frac{1}{2\pi i \tau}\int_0^\pi \frac{[x+(x^2-1)^{1/2}\cos\gamma]^2 - 1}{[x+(x^2-1)^{1/2}\cos\gamma]^{3/2+i\tau}} \, d\gamma. \tag{A11}$$

In assessing convergence of the integrals appearing in the solution, the asymptotic behaviors of the functions are needed. Using Eq. (A5) in conjunction with Eqs. (4) and (6) of [30] leads to:

$$\lim_{\tau\to\infty} C_{1/2+i\tau}^{-1/2}(x) = \frac{1}{\tau^2 + 1/4}\sqrt{\frac{2}{\pi\tau}}\left\{-\frac{x}{(x^2-1)^{1/4}}\sin[(\cosh^{-1}x)\tau + \pi/4]\right.$$

$$\left. + \tau(x^2-1)^{1/4}\cos[(\cosh^{-1}x)\tau + \pi/4]\right\}, \tag{A12}$$

$$\lim_{\alpha\to\infty} C_{1/2+i\tau}^{-1/2}(x) = -\sqrt{x}\sqrt{\frac{2}{\pi}}\frac{|A|}{\tau^2+1/4}\left\{\frac{\cos[(\cosh^{-1}x)\tau + \text{Arg}(A)]}{2}\right.$$

$$\left. + \tau\sin[(\cosh^{-1}x)\tau + \text{Arg}(A)]\right\}, \tag{A13}$$

where,

$$A(\tau) = \frac{\Gamma(i\tau)}{\Gamma(i\tau + 1/2)}, \tag{A14}$$

in which $\Gamma(z)$ is the Gamma function [25]. From the representation of the Gegenbauer functions given by Eq. (A11) along with its asymptotic behavior from Eq. (A13), it is seen that the transform will exist provided that $\lim_{z \to \infty}[f(x)/\sqrt{x}]$ is finite and $f(1) = 0$.

Finally, it is noted that the transform presented here, may be related to the Mehler-Fock transform of zero order [28]. Using Eq. (A5) in Eq. (A7) and then integrating by parts results in:

$$\int_1^\infty f(x)(x^2-1)^{-1} C_{1/2+i\tau}^{-1/2}(x)\, dx = (\tau^2 + 1/4)^{-1} \left\{ [f(x) P_{-1/2+i\tau}(x)]_1^\infty \right.$$

$$\left. - \int_1^\infty f'(x) P_{-1/2+i\tau}(x)\, dx \right\}. \tag{A15}$$

The last term is the Mehler-Fock transform of zero order of the function $f'(x)$. The transforms required in the present study were evaluated from the above equation by using the Mehler-Fock transforms tabulated in [28].